\documentclass[12pt]{iopart}

\usepackage{graphicx}
\usepackage{soul}
\usepackage{todonotes}

\begin{document}

\title[Origins of Life: A Problem for Physics]{Origins of Life: A Problem for Physics}

\author{Sara Imari Walker}

\address{School of Earth and Space Exploration and Beyond Center for Fundamental Concepts in Science, Arizona State University, Tempe AZ USA; Blue Marble Space Institute of Science, Seattle WA USA}
\ead{sara.i.walker@asu.edu}
\begin{abstract}
The origins of life stands among the great open scientific questions of our time. While a number of proposals exist for possible starting points in the pathway from non-living to living matter, these have so far not achieved states of complexity that are anywhere near that of even the simplest living systems. A key challenge is identifying the properties of living matter that might distinguish living and non-living physical systems such that we might build new life in the lab. This review is geared towards covering major viewpoints on the origin of life for those new to the origin of life field, with a forward look towards considering what it might take for a physical theory that {\it universally} explains the phenomenon of life to arise from the seemingly disconnected array of ideas proposed thus far. The hope is that a theory akin to our other theories in fundamental physics might one day emerge to explain the phenomenon of life, and in turn finally permit solving its origins.
\end{abstract}

\maketitle

\tableofcontents

\section{Introduction}

The origin of life is among the greatest open problems in science -- {\it How is it that life can emerge from non-living matter?} An answer is critical for understanding our own origins, for identifying the most promising targets in the search for life on other worlds, and for synthesizing new life in the lab. Despite the significance of the problem, we currently have few scientific windows into its resolution. Historically, the origin of life has been viewed as a problem for chemistry \cite{LM1996, Orgel1998, Cleaves2011}. Most research has centered on the biochemistry of life as we know it on Earth today, after over $3.5$ billion years of biological evolution. This includes producing RNA, peptides, lipids or components of extant metabolisms from prebiotically plausible conditions. However, given the uncertainties in our knowledge of the earliest life, or of its host environment, a focus solely on the historical sequence of events leading to the emergence of life may be hindering progress: it is unknown what features of life are likely to be {\it universal}, characteristic not only of life as we know it now, but also the simpler life-forms that first emerged here on Earth and life as it might exist elsewhere. For example, we do not yet know if life could exist with a different chemistry than natural biochemistry, or whether life might have started this way, even on Earth. Due to these limitations, there has been increased interest in studying the origin of life not strictly as a problem for chemistry (or biology), but as a problem for physics \cite{GW2011, WD2013, SM1982, SM2016}, with potentially deep implications for our most fundamental understanding of the natural world. 

In order for physics to be able to address important open questions regarding universal features of the living state -- akin to ``laws of life'' -- it seems that our view of the origin of life must too evolve. For one, understanding the origin and nature of life may not be separable problems as often thought \cite{Szostak2012}: windows into one will likely provide important insights into the other. Much research on the origin of life has tacitly averted addressing the question {\it ``what is life?''}, by focusing on synthesizing the chemical constituents of known life or by attempting to build a simple chemical replicator capable of Darwinian evolution. Significant progress has been made in understanding minimal chemical systems with these properties. However, many researchers still regard defining life as an intractable problem, which is a nonconstructive digression from solving its origins. The problem of course is that what is needed is not a definition for life, but a theory of life from which useful criteria for evaluating competing models for the origin of life might emerge. Currently no such criteria exist. The simple models for the emergence of life produced in the lab and {\it in silico} thus far are a far cry from the rich dynamics exhibited by living systems. It is not that our current theories of physics cannot accommodate life, certainly living processes do not violate any of the known laws of physics, but those same laws do not {\it explain} the phenomenon of life. We are however not without hope that a theory for the living state is within reach. 

This review is geared towards covering major viewpoints on the origin of life, with a forward look towards considering what it might take for a physical theory that {\it universally} explains life to emerge from the seemingly disconnected array of ideas proposed thus far. While there is a large body of work studying fundamental properties of life from the perspective of physics (see {\it e.g.} Bialek \cite{Bialek2012} and and Kaneko \cite{Kaneko2006} for physical and dynamical systems perspectives on the subject, respectively),  this review is primarily intended for physicists interesting in pursuing origins questions to gain a broad overview of recent thinking in the field as well as future directions the field might take. I therefore primarily review approaches explicitly focused on the origins life, and only briefly touch on the broader topic of fundamental physics of life. The spirit of this review is intended to interest a broad readership as new thinking from diverse perspectives will be necessary for uncovering principles of living organization akin to our other theories in fundamental physics. The hope is that we may one day explain the phenomenon of life, and in turn finally permit solving the origins of life.

\section{Knowns and unknowns in solving the origin of life}

Our uncertainty about the origin of life can be summed up succinctly as our ignorance in calculating the probability, ${\cal P}_{\rm life}$, for matter to transition from the non-living to living state. This is an important parameter to know, not just for understanding life on Earth, but also for estimating the distribution of life in the universe. It therefore ranks with other parameters in cosmology as defining important features of our universe. However, at present there exist very few constraints on ${\cal P}_{\rm life}$ since we know of only one planet that is inhabited: our own.

Of the major historical unknowns in constraining the origin of life on our planet -- the {\it where}, {\it when} and {\it how} life first emerged -- we have the least certainty in {\it how} and {\it where}. Only the timescale for life's emergence seems reasonably constrained. Fossil evidence for cellular life exists early in the geological record, dating back at least 3.5 billion years \cite{Schopf} \footnote{It is important to note that the earlier the sample, the more difficult it is to definitively identify biogenic origin see {\it e.g.}  \cite{Brasier2006,vZLA2002, Dodd2017}, so evidence for life earlier than $\sim 3$ Bya is often subject to intense debate.}. These earliest fossils are examples of stromatolites, mineral mounds created by microbial communities, which are still found today in shallow, hypersaline waters, such as in Shark Bay in Australia. Despite this antiquity, stromatolite communities are not representative of the first life; formation of stromatolite mineral deposits requires cells with advanced biomolecular machinery (inclusive of the capacity for photosynthesis as cyanobacteria are important contributors to the formation of stromatolites): indicative of cellular life with a complexity comparable to organisms alive today. The early appearance of stromatolites in the fossil record therefore places relatively tight constraints on the timescale for the origin of life: life had to emerge early for evolution to lead to such complex communities. Conditions became habitable on Earth approximately $4$ billion years ago (Bya): life therefore had to emerge and evolve 'modern' cellular complexity within a window of just a few hundred million years. More speculatively, there also remains open the possibility that life did not emerge on Earth (the hypothesis of panspermia), and instead was delivered to Earth by impacts -- for example, life could have originated on Mars and been delivered to Earth. Panspermia potentially increases the window of time within which the origin of life might have occurred \cite{Davies2003}, but adds its own complications in that it places the origin of life in an unknown environment that is even less constrained than what little we know about the environments on the early Earth.

Assuming the simplest case, that life did indeed emerge on Earth, we can reasonably assume that the early evidence for complex cellular life suggests that the origin of life was a rapid event (in a geological sense), occurring nearly as soon as conditions were favorable. A corollary to this argument is that since life appeared so rapidly on Earth, the origin of life must be a relatively common event on Earth-like worlds \cite{LD2002}. One might therefore conclude that ${\cal P}_{\rm life} \rightarrow 1$, at least for Earth-like planets in the habitable zone of their parent star (as is often done in optimistic estimates in astrobiological searches for life and intelligence in other planetary systems). This argument, combined with increasingly frequent discoveries of ``Earth-like'' worlds \cite{Borucki2012,Borucki2013}, has lead many astrobiologists to be optimistic that life should be common in the universe. If true, life could be detected in the atmosphere of an exoplanet with the next generation of telescopes, such as the James Webb Space Telescope \cite{Cowan}. If we are lucky, and do discover life readily in the next decade, we will have new constraints on the planetary contexts that allow for life, a discovery which in turn might help inform our understanding of the origins of life.

\subsection{One planet, one sample: The significance of anthropic bias} \label{sec:anthropic}
In the absence of such a discovery, with current empirical data it is impossible to determine with any certainty whether life is common or rare. The challenge arises because we have only one sample of life on which to inform estimates of ${\cal P}_{\rm life}$. All known life on Earth shares a common ancestry, descending from a last universal common ancestor (LUCA) that lived on Earth more than $3.5$ Bya. LUCA in reality should be thought of as community of organisms undergoing rapid horizontal exchange of genetic information, rather than an individual cell or species as the name might imply \cite{Woese1998}. Evidence for this common ancestry derives from phylogenetic reconstruction of the history of life on Earth, as inferred from modern organisms \cite{Theobald2010}\footnote{It should be noted that while it is clear that the tree of life has a common root, it is not as clear where to place the root \cite{Doolittle}).}. Another line of evidence corroborating phylogenetic reconstruction is the existence of universal features of biochemistry shared by all organisms on Earth (such as DNA and RNA which allow phylogenetic histories to be generated in the first place). Examples of this universal biochemical toolkit are included in Table \ref{Table:univ}. 
Among these are the utilization of deoxyribonucleic acid (DNA) for storing genetic information, ribonucleic acid (RNA) for transcribing that information and proteins for performing biochemical catalysis. In short, we have direct evidence for only one origins event \footnote{Even though the origins of life may in reality be a composite many of ``origins'' and symbiosis events in earlier evolution (see {\it e.g.} \cite{Dyson1982}), we do not have direct evidence for this and it does not change the nature of the anthropic selection arguments presented here in any case - for that, all that matters is there is a universal ancestry for {\it extant} life.}. 

\begin{table}
\centering
\begin{tabular}{  p{11cm} } 
{\bf Universality in the chemistry of known life} \\
\hline 
\hline
  DNA as the genetic material \\ \hline 
  A genetic code composed of three-nucleotide codons \\ \hline 
  RNA as the intermediate in expressing genetic information \\ \hline 
  Translation machinery including ribosomes and tRNAs \\ \hline 
  Proteins as biochemical catalysts \\ \hline 
  Homochirality (L-isomers of amino acids, D-isomers of sugars)\\ \hline 
  ATP as an energy intermediate\\ \hline 
  Lipid bilayer membrane \\  \hline 
  \hline
\end{tabular}
\caption{Some of the universal features common to the biochemistry of all known life on Earth.} \label{Table:univ}
\end{table}

Because we are constrained by a single biochemical sample of life, despite our best attempts at logical arguments to the contrary, we cannot say whether the origin of life is easy or hard. That is, {\it the probability ${\cal P}_{\rm life}$ is unconstrained}. Carter was the first to quantify this, demonstrating via Bayesian reasoning that careful analysis of the observational facts (that life arose once, and seems to have done so rapidly once conditions were favorable) is equally consistent with life being very common and also with it being exceptionally rare \cite{Carter}. Spiegel and Turner later did a more formal analysis and arrived at the same conclusion \cite{ST2012}. Here I follow the reasoning of Carter's original logic, which relies on Bayes theorem, and relays the essential points of the argument (see \cite{ST2012} for a more technical assessment). Via Bayes theorem we have: 
\begin{eqnarray}
P(t|d) = \frac{P(t|d) P(t)}{P(d)} ~.
\end{eqnarray}
This is a statement about conditional probabilities, and follows from considering the joint probability of two statements $t$ and $d$ both being true, $P(t, d)$.  This may be expanded as $P(t, d) = P(t|d) P(d)$ where $P(d)$ is the probability of $d$ being true and $P(t|d)$ is the probability of $t$ being true contingent on the fact that $d$ is also true (this is called a conditional probability, see {\it e.g.} \cite{Jaynes}). For purposes of discussion of Carter's argument, we may consider $t$ to represent a {\it theory} and $d$ observational {\it data}. Comparing two alternative theories, {\it e.g.} that life is a common occurrence (denoted here by $t_c$ for common) or that life is very rare (denoted here by $t_r$ for rare) yields:
\begin{eqnarray} \label{Bayes:hyp}
\frac{P(t_c|d)}{P(t_r|d)} = \frac{P(d|t_c) P(t_c)}{P(d|t_r) P(t_r)}~.
\end{eqnarray}
Carter's point is this: the effects of experimental bias and observational selection must be taken into account when computing the values for the likelihood of the observed data ($P(d|t_c)$ and $P(d|t_r)$) on the right-hand side of the above equation. These represent probabilities based on {\it current knowledge}, and are often incomplete -- that is, they are not {\it ab initio} probabilities derived from a fundamental theory and do not represent absolute states of knowledge. They are thus subject to anthropic selection effects. 

If for example, we discovered a second sample of life -- either on another planetary body, or a second sample here on Earth \cite{shadow}, then we might have reasonable confidence that $P(d|t_c)/P(d|t_r) \gg 1$. That is, we would have sufficient data to support the induction that life is common over the alternative hypothesis that it is rare (our data would be more consistent with the hypothesis $t_c$ than with $t_r$). Of course no such discovery has yet been made. Under the constraint of a single sample, we can only operate under post-selection (anthropic selection) on the likelihood of our observations. That is, one of the {\it a priori} conditions we must account for is that the first planet available for us to investigate must include the prior occurrence of life. This post-selection, under the constraint of a single data point, collapses the probabilities on the right-hand side of Eq. \ref{Bayes:hyp} to $P(d|t_c)/P(d|t_r) = 1$, since both hypotheses are equally consistent with current data. We cannot distinguish between the hypothesis that life is common (hypothesis $t_c$) or that it is rare (hypothesis $t_r$): both give equally consistent data based on current observational evidence. Stated more simply, if life is exponentially rare on Earth-like worlds (or any worlds) such that we should expect it to occur only once in a Hubble volume\footnote{The Hubble volume is the region of the universe we are in causal contact with, {\it i.e.}, the observable part.}, we should of course expect to find ourselves on that planet. Carter further goes on to argue that our observational evidence is more consistent with the rare hypothesis, due to the coincidence in time scales between our existence today, and the habitability window for Earth (which will terminate in $\sim 800$ million years when our Sun leaves the main sequence), a coincidence made more probable if life is in fact very rare (see \cite{Carter} for discussion). 

\begin{figure}
\centering
\includegraphics[width=5in]{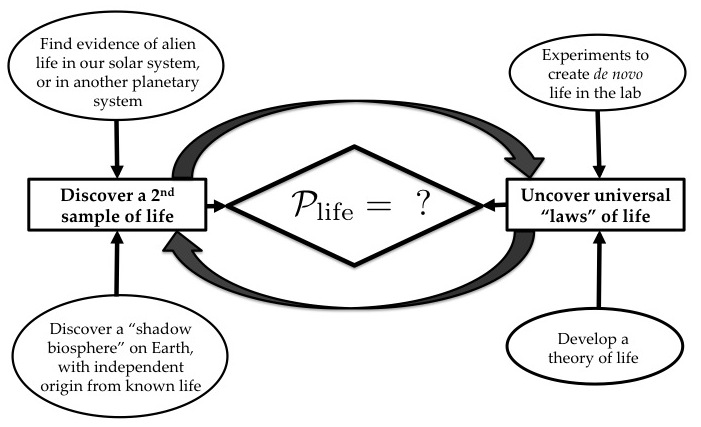}
\caption{Possible constraints on ${\cal P}_{\rm life}$, the probability that life emerges from non-living matter.} \label{fig:plife}
\end{figure}

\subsection{Two paths to a solution}

Without a reasonable estimate for the likelihood of either hypothesis (that the origin of life is hard or easy) we have very few constraints on ${\cal P}_{\rm life}$. The {\it most} we can definitively say is that ${\cal P}_{\rm life} \subset (0,1]$, where we can exclude the possibility ${\cal P}_{\rm life} \equiv 0$ since we are here to ask about it. Thus, in order to address the problem at hand we need to find better ways to constrain ${\cal P}_{\rm life}$. There are two ways out of Carter's argument, as illustrated in Fig. \ref{fig:plife}: either we observationally discover a second sample of life (thus increasing the likelihood $P(t_c|d)$ over that of $P(t_r|d)$), or we uncover the mechanisms governing the transition from non-living to living matter. The latter would allow estimation of an unbiased {\it ab initio} probability directly from theory. 

It does not matter if alien life is discovered extraterrestrially or if it is discovered here on Earth, so long as it has an independent origin. This has led to proposals that our best chance of constraining ${\cal P}_{\rm life}$ is to discover a second sample of life here on Earth, a so-called ``shadow biosphere'' that has so far eluded detection \cite{shadow}. Two factors stack in favor of this proposal over that of looking for alien life elsewhere: (1) we know life has arisen on Earth at least once and (2) it is technically and financially easier to look for unknown life here on Earth than to develop missions to detect life in other planetary environments in the solar system or on exoplanets. One popular hypothesis is that life could exist with a different chirality to that of known life (see Section \ref{Sec:Chiral}). Other ideas include looking for life utilizing alternative amino acids, more exotic chemical constituents, or even other elements \cite{shadow}. The challenge is that if biochemically `alien' life exists, it has so-far defied detection, and the weirder it is the harder it will be to discover. When looking for new life-forms on Earth, the presence of DNA is the most indicative and widely used biomarker. An example are efforts to comb the world's oceans to uncover patterns in biodiversity and potentially new life -- detected precisely by the presence of nucleic acids \cite{globalocean}. Any non-DNA based life goes unnoticed, being impossible to detect with current methods. Discovering ``shadow life'' is not without precedent: archaea were once entirely unknown, until finally discovered as the third domain of life by 16S RNA sequencing (almost 300 years after the discovery of bacteria) \cite{3rddomain}. A better idea of what distinguishes life from non-life is necessary.  

There are also advantages to searching for a second sample of life beyond Earth. These include the potential to discover life very different than what is known. Importantly, the sheer number of potential environments and locals provide many locales to look. In the solar system alone there are at least five targets that are regularly discussed in astrobiology, these include: Mars, Europa, Enceladus and perhaps more speculatively Titan and Venus. Life existing on an icy moon, such as Europa, Enceladus or Titan for example, would likely be very different from life on Earth -- allowing the possibility for new insights into universal properties of life. Beyond our solar system, exoplanet research is now poised to provide unprecedented opportunity to discover alien life. With the discoveries of Kepler \cite{Batalha} and other searches, hundreds of worlds have been discovered. So far, much of the discussion has been focused on 'Earth-like' worlds, {\it e.g.}, those orbiting K (yellow) stars like our own Sun, that are located in the 'habitable zone' where liquid water might exist on the surface. However, new planet-types, not represented in our own solar system, have also been discovered including 'super-Earths' with masses of five Earth masses ($M_\oplus$) and 'mini-Neptunes' with masses up to about $\sim10 M_\oplus$ (Uranus and Neptune have masses of $\sim 14.5$ and $17$ $M_\oplus$ respectively). Due to the ambiguity associated with many putative biosignatures REF and the very real potential for detecting false positives for life \cite{Schwieterman}, the state of the art in exoplanet life detection is currently focused on identifying a suite of biosignature gases, combined with deep knowledge of planetary context. Recent proposals include cataloging all small molecule volatiles that could potentially exist in a planetary atmosphere, inclusive of their abiotic and biological sources and sinks \cite{Seager}. This is a monumental task and will likely take decades, a timescale on par with likely mission timescales. 

Avenues for developing new approaches could come from the interface of the bold arrows in Fig. \ref{fig:plife}, that is from feedback between communities searching for life on exoplanets and those in the complex systems and physics communities studying the fundamental properties of living matter. As just one example, several studies have hinted that the network topology of atmospheric chemical reaction networks might be different on inhabited worlds, and therefore could be an atmospheric biosignature \cite{Sole, Jolley}. In particular, Earth's atmospheric reaction network is highly heterogeneous with a few chemical species that act as hubs and are involved in many reactions, whereas many other chemical species are involved in only a few. Generating a network representation where chemical species are nodes and edges connect species involved in the same reaction reveals a power-law like behavior in the global topology of Earth's atmospheric reaction network \cite{Sole}. This is significant as many biological systems, including metabolic chemical reaction networks display similar topological properties \cite{Jeong}. Other planetary atmospheres, such as those of Mars, Venus and Titan, are more homogenous in their network organization, with global topology more consistent with random networks \cite{Sole}. Static reaction graphs, based on reaction lists for equilibrium solutions of planetary atmosphere models are a first step, but more rigorous work remains to be done to understand the network structure of planetary atmospheres, how this varies as a function of dynamics and kinetics, and whether or not those of living worlds are statistically distinct from non-living worlds. Feedback between the exoplanet and origins of life communities need not go in one direction either: understanding planetary atmospheres of inhabited and uninhabited worlds could also help inform our understanding of chemical organization within living systems by providing null chemical models for their large scale organization \cite{Holme}. 

New techniques for how we approach the problem of detecting alien life could also help inform our understanding of its universal properties. So far, the majority of research into exoplanet biosignatures has focused on detecting life in atmospheric gases on an individual target planet. However, if life is rare (or at least not so common that it exists on every Earth-like world), or if life can be very different than known life, it may be that this targeted approach is not the right method. This could be either because we choose the wrong target, the wrong biosignature, or simply that we cannot resolve life above the background geochemistry. The latter is a real possibility given that planetary evolution models are stochastic and predict the properties of planets probabilistically, yielding uncertainties in our ability to pin down the abiotic background of a given target. Alternative approaches based on sampling large statistical ensembles of exoplanet atmospheres could avert this issue, as models can predict the distributions of properties of exoplanets with similar properties. Constrain ${\cal P}_{\rm life}$ with a sufficiently large statistical ensemble of observed exoplanet spectra would also place important constraints on the environments where life might be likely to arise and thus inform our understanding of its origins. For the remainder of this review, I will be less concerned with the approaches on the left of Fig. \ref{fig:plife} and will primarily address those on the right. Physics potentially has the most to contribute to the problem of solving life's origin by uncovering universal ``laws'' governing biological organization. However, the future of both fields may dependent on tight integration between communities of researchers studying origins and those studying biosignatures.

\section{Defining life and solving its origin are not different problems}

We do not yet know if there are universal ``laws'' that underlie biological organization. However, if the trend goes as in other areas of physics we might expect that we will one day uncover them. 
In the absence of a theory explaining living matter, scientific approaches to the origin of life must either explicitly or implicitly adopt a working definition to make progress. Thus, there are nearly as many theories for life's origins as there are definitions for life (perhaps more so). These two problems -- defining life and solving its origin -- cannot be readily disentangled \cite{ClelandChyba}. 

One might, for example, take a purely substrate-level definition for life and conjecture that life is defined by its constituent molecules, including amino acids, RNA, DNA, lipids {\it etc.} as found in extant life. It then follows that the problem of life's origin should reduce to identifying how the building blocks of life might be synthesized under abiotic conditions (which as it turns out is not-so-easy). This approach has dominated much of the research into life's origins since the 1920's when Oparin and Haldane first proposed the ``primordial soup'' hypothesis, which posits that life arose in a reducing environment that abiotically synthesized simple organic compounds, concentrated them, and gradually complexified toward more complex chemistries and eventually life \cite{Bernal1967}. In 1953 Miller demonstrated that organic molecules, including amino acids, could be synthesized in a simple spark-discharge experiment under reducing conditions \cite{Miller}. At the time, there was such optimism that the origin of life problem would soon be solved that there was some expectation that life would crawl out of a Miller-Urey experiment within a few years. This has not yet happened, and there seem to be continually re-newed estimates that artificial or synthetic life is just a few years away. This suggests a radical re-think of the problem of origins may be necessary \cite{CW2016}. 

One challenge is that the origin of life on Earth happened in the remote history of our planet. Much of the record of the first living systems and the environment they occupied has been erased by subsequent evolution of the biosphere and geosphere. It is therefore entirely possible that the first living systems looked very different than those today in terms of their chemical composition. There is increasing evidence in support of the view that early life could have had a very different chemistry than that articulated in Table \ref{Table:univ}, a view first suggested decades ago by Cairns-Smith in his model of clay life \cite{CS1982}. Cairns-Smith envisaged the first life to be instantiated in inorganic clays that could template-replicate, which transitioned via ``genetic takeover'' to the more familiar organic genetic polymers of extant biochemistry. The core idea was to move beyond a substrate-based definition for life to a definition that depends on other factors: namely, in Cairns-Smith's case, the replication of heritable information. While in details this proposal remains speculative, the core idea is productive in challenging our pre-conceived notions of what chemistries could potentially be ``biological'' \cite{report}. More modern approaches to the problem are indeed revealing that life is likely not exclusive to its known substrates. For example, synthetic biology has already demonstrated that components of the chemistry of life can be expanded to molecules that are not biological in origin, such as unnatural base pairs \cite{expandgenetic} or alternative nucleic acids \cite{Pinheiro2012}. Addressing the origin of life problem therefore requires more than answering the historical question of how life arose on Earth - it requires understanding universal features of the transition from the non-living to living state, even if we do not know (and may never know) the exact chemical nature of the first life on Earth. 

Although we do not know yet what these universal features might be, different working definitions for life suggest different routes for uncovering the principles underlying the transition from non-living to living matter. Many of the most common definitions for life may be found compiled by Trifonov \cite{Trifonov}. However, definitions should emerge from theories, not the converse \cite{ClelandChyba}. 

\section{Life from (bio)chemistry} \label{Sec:chem}

Chemical definitions for life take on many forms, focusing either explicitly on the chemistry of extant life (as discussed in the previous section) or the organizational principles of biochemistry. Traditionally, these approaches have fallen into two camps ``genetics-first'' and ``metabolism-first'': however recent progress in the field is starting to merge these two perspectives into a more cohesive view of the processes that drove the emergence of life. 

\subsection{Life as genetics}

The earliest life-forms on Earth could have been very different than modern life biochemically. It is not known just how different core features of biochemistry could be from those in Table \ref{Table:univ} and still be viable. No matter the chemistry, many agree life should be capable of {\it genetic heredity}. Following Adami, one might take an abstract (and therefore more universal) definition, and regard life as ``information that copies itself'' \cite{AdamiLabar}. The information can be encoded in a genetic molecule (or a clay as Cairns-Smith proposed) and this information is propagated from generation to generation (is copied). In a less extreme version of ``genetic takeover'' than that proposed by Cairns-Smith, many have conjectured that prior to the evolution of a deoxyribonucleic acid (DNA) genome, putative ``ribo-organisms'' could have utilized ribonucleic acid (RNA) as their sole genetic material (a more recent variant not within the ``genetics-first'' framework even proposes the notion of a ribo-film as an early stage of life, see {\it e.g.} \cite{BarossMartin}). Such life-forms would be very different than an extant life today, and would look quite alien to us in terms of their biochemistry. Defining life as `genetics' allows exploring the possibilities for such putative ``alien life''. 

\subsubsection{The ``RNA World''.}
The hypothesis that ancestral genetic systems were based on solely on RNA is called the ``RNA world'' \cite{Gilbert1986}. In modern interpretations this may be considered as an umbrella term for a number of hypotheses about early life that are unified by the view that RNA preceded DNA as the primary molecule of information storage and heredity in the evolution of genetic systems on Earth. There are alternative takes what this implies for the first living systems. One end of the spectrum advocates the view that RNA was the first living thing, arising directly from abiotic sources of organics on early Earth (and is therefore implicitly focused primarily on the substrate of life as being its defining feature). This view has been challenging to substantiate given the difficulty of synthesizing RNA under prebiotically plausible conditions \cite{Shapiro1988, LM1996}. Recent progress however has been made in synthesizing both purine and pyridine ribonucleotides in so-called ``one-shot'', or multi-component, experiments by the Sutherland lab \cite{PSS2010,PGS2009}. There is also the challenge that RNA is unstable in water (dubbed the ``probability paradox'' by Benner \cite{Benner2014}), which means that even if we could find conditions for synthesizing RNA polymers, they degrade rapidly (although this also holds potential for being an evolutionary advantage speeding up the search time to find functional biopolymers, see \cite{Mathis2016, WGH2012}). The RNA-first view has thus far primarily held sway because of its simplicity, as well as the relative ease of studying RNA systems in the laboratory as empirical models for exploring molecular evolution \cite{PBC2015, HL2015}, however new approaches are broadening this view. 

\subsubsection{Alternative Genetic Polymers.}
With recent advances in systems chemistry, alternative chemical models are becoming increasingly tractable. These include models for both the actual historical sequence of events of the origin of life that could have occurred on early Earth \cite{MKA1999,Hud2013}, and synthetic systems to test the principles of biogenesis in the lab that do not include natural biomolecules \cite{report, Sadownik2016,Cronin2012,ZL2997, Otto2015}. There is also increasing evidence in support of the hypothesis that the chemical nature of the first living systems on Earth could have been very different than that of modern life. We have ample supporting evidence that life may have undergone a ``hardware'' upgrade at least once in the evolution from RNA to DNA genomes (see {\it e.g.}, \cite{Leu2011}). An analogy to ``my grandfather's axe'' has been made for the process of chemical evolution of other genetic polymers that could have preceded even RNA \cite{Hud2013}. An axe can have both the handle and head replaced and nonetheless retain its functionality. Likewise, if genetic systems can be shown to retain functionality while swapping out their individual components (the nucleobases, ribose, and phosphate) over time in an evolutionary succession of genetic polymers, it opens the possibility that the first genetic polymers could have been very different than RNA or DNA \footnote{An interesting philosophical question is whether the grandson's or granddaughter's axe is still the same axe, that is would the original genetic system represent the same sample of life (given common descent) or different (given different core chemistry)?}. 

There have been a number of studies conducted exploring the chemical etiology of nucleic acid structures \cite{Eschenmoser1999}, aimed at systematically exploring the landscape of possible genetic polymers related chemically to DNA and RNA. Thus has been born a plethora of ``pre-RNA world'' hypotheses, suggesting that just as DNA may have replaced RNA, other genetic systems may have preceded RNA. Additional support for this view comes from the fact that alternative nucleic acids or XNAs, such as peptide nucleic acid (PNA) \cite{NLM2000} or threose nucleic acid (TNA) \cite{Orgel2000}, are easier to synthesize under prebiotic conditions than RNA. Furthermore, while the chemical universe of nucleic acid structures is immense, constraints can nonetheless be imposed on which nucleic acids are viable precursors to RNA as only certain combinations of nucleic acids will mediate {\it information transfer}. For example, it has been shown that RNA can exchange information (template) with glycerol nucleic acid (GNA) or TNA, but TNA and GNA cannot directly exchange information with one-another \cite{Chaput2007}. If either GNA  or TNA was an immediate predecessor to RNA in the evolution of genetic systems, than sequentially the other polymer species is excluded. 

Additional support for the pre-RNA world hypothesis comes from an entirely different direction -- synthetic biology. Six XNAs have recently been shown to be viable functioning polymers, capable of Darwinian evolution, aptamer activity \cite{Pinheiro2012} and even catalysis \cite{Taylor2015}. Other refinements on genetic systems are possible. Another proposed variant of early genetic polymers is RNA with a mixed backbone linkages \cite{EPS2013}. The heritable information in nucleic acids consists of the sequence of nucleobases. Mixed backbone polymers therefore pose an intriguing problem for understanding mechanisms of functional heredity since the backbone structure constitutes non-heritable information, yet the backbone structure contributes significantly to folding and thus function. While the chemical details of many of the alternative nucleic acid architectures discussed herein remain a subject of intense laboratory investigation, less work has been done to understand important aspects of the dynamics governing both non-heritable aspects of functional information or the dynamics of information transfer through ``genetic takeover'' events. Thus, for example, while in principle RNA can transfer information (template) with TNA and GNA it is unclear how the chemical properties of these systems might yield differences in the evolutionary transitions between genetic systems, such as in the fidelity and accuracy of information transfer. A necessary step forward is to build viable {\it evolutionary} models (empirical or theoretical) that definitively demonstrate life can survive such a dramatic change in the chemical nature of its component parts. Computational models, for example, could constrain when information transfer events between different genetic systems should be expected to occur, and in what direction -- {\it i.e.}, to address the question when can information copy itself? From the perspective of physics, models must be developed for understanding how functional information propagates and is preserved across distinct physical media.

\subsubsection{Limitations of ``Genetics-first'' Models.} \label{sec:gen_lim}
A final point on genetics-first models is the implicit definition of life assumed: that life is defined by heritable replication of genetic information and selection on that information. This view adopts a Darwinian criterion for life, with the commonly assumed corollary that no matter how simple the system, if it is capable of Darwinian evolution `life' will eventually arise. This solution to the origin of life problem thereby reduces to one of identifying a primordial polymer that could form abiotically and jump-start the Darwinian evolutionary process. This makes sense from the perspective of extrapolating backward from biology, but perhaps less-so when the constraints of physics and chemistry are taken into account.  A commonly cited constraint is the error-threshold (see {\it e.g.} \cite{BE2005}), which places limits on the error-rate permitted for the information content of genetic systems to be heritable \cite{MSS1997}. However, the concept of an error-threshold as proposed by Eigen and Schuster assumes that all molecules in a population can replicate: in reality most random sequences are inert and cannot replicate. We should therefore not expect the majority of random sequences to be capable of self-replication. In fitness landscapes where there exist lethal mutations, no error threshold is observed \cite{wagner1993}. A further point is that the information that contributes to fitness can be encoded in many different ways, leading to `neutrality selection' whereby selection acts to increase the probability that mutations are selectively neutral. The relevant equation for the fraction of a replicating population that dies each generation due to deleterious mutations (the mutational load) is  \cite{OAC2003}:
\begin{eqnarray}
L \sim 1 - e^{-Rl (1-v)}
\end{eqnarray}
Where $R$ is the error-rate per monomer copied, $l$ is the length of the sequence, and $v$ is the fraction of mutations that are not deleterious (are selectively neutral). A population can reduce its mutational load at constant mutation rate by increasing the neutrality $v$. Studies of real ribozymes demonstrate selectively neutral landscapes, leading to a relaxed error threshold when one accounts for selection on the phenotype of the molecule \cite{KSS2005}. In early chemical systems, neutrality selection could for example be accomplished via redundancy, permitting early replicators to produce functional copies while maintaining high population diversity even in the face of high mutation rates. Feedback between the composition of replicators and their environment could also act to regulate selection dynamics to maintain heritability \cite{VWL2012, Mathis2016}. The challenge of genetics-first is to determine the likelihood for discovering replicating polymers by chance, which depends on the environment. The probability of the emergence of self-replicating polymers is currently unknown. Spontaneous synthesis of genetic polymers remains a major technical challenge for prebiotic chemistry and casts some doubt on the utility of `replication-first'. Even in the context of the RNA-world, the first role of RNA may not have been templated polymerization, RNA is known to undergo recombination and can spontaneously form networks \cite{NVetal}-- perhaps the earliest roles for RNA were not strictly confined to a `DNA-like' role. 

Another problematic aspect is the assumption that Darwinian evolution invariably leads to greater complexity. 
A now classic example in the molecular evolution literature is Spiegelman's monster, a $4500$ nucleotides RNA virus evolved {\it in vitro}, which through competitive selection on replication evolved to be as short as $218$ bases (the ``monster'') \cite{Spiegelman}. Spiegelman's monster is an example of `compression selection' \cite{OAC2003}, whereby information is lost from a genome when it is no longer relevant to its fitness. This occurs, for example, when a replicator's environment becomes more simple. A trend of increasing complexity requires complex, information-rich environments, not just the capacity for self-replication. Thus, the environment and selective pressures must be included in any discussion of the origins of life that aims to account for how simple chemical systems could increase in complexity over time \cite{peptideengine}. There is the further issue that all else being equal, thermodynamics may actually favor simpler replicators \cite{England2013} (more on this in Section \ref{sec:SMrep}). The genetics-first paradigm, while permitting a pathway for `life-like' evolution, still leaves open the questions of identifying the environments that increase the probability for the spontaneous emergence of self-replicators or the probability that they subsequently evolve towards states of {\it increasing} complexity over time. In particular, it remains to be identified whether the environments that favor the spontaneous emergence of replicators also favor their evolvability. 


\subsection{Life as cooperative networks} \label{sec:auto}

There exist many working definitions for life that differ from the Darwinian view, each leading to alternative hypotheses to the genetics-first picture for life's emergence. A primary competitor to a genetic replication-based scenario is that life first emerged as set of molecules that could {\it collectively} reproduce, that is, as an autocatalytic set. Autocatalysis occurs when the product of a reaction is a catalyst for that same reaction or a coupled reaction. {\it Autocatalytic  sets} arise when a group of molecular species forms a set of catalysts where reaction(s) producing each species in the set are catalyzed by at least one other species within the set \cite{Kauffman1986, Kauffman1993, HHS2010}. Thus, a focus on autocatalytic networks adopts a definition of life as a self-organized phenomenon based on the principle of collective reproduction. 

\subsubsection{Autocatalytic Sets.}
While originally proposed by Kauffman as a model for the origin of life several decades ago \cite{Kauffman1986}, it was only very recently that the idea of an autocatalytic set has been mathematically formalized in Reflexively Autocatalytic Food-generated (RAF) theory, as developed by Hordijk and Steel \cite{HS2004} and adopted as useful tool for thinking about the organization of chemical networks within the systems chemistry community \cite{AJYG2004, SvK1994,LJ2009, Vaidya2012, HS2013}. The definition of a RAF considers a network of catalyzed chemical reactions, a (sub)set $R$ of which are called:
\begin{itemize}
\item {\bf Reflexively autocatalytic} (RA) if every reaction in $R$ is catalyzed by at least one molecule involved in any of the reactions in $R$;
\item {\bf F-generated} (F) if every reactant in $R$ can be constructed from a small food set $F$ by successive applications of reactions from $R$;
\item Reflexively autocatalytic and F-generated (RAF) if it is both RA and F.
\end{itemize}

\begin{figure}
\centering
\includegraphics[width=3in]{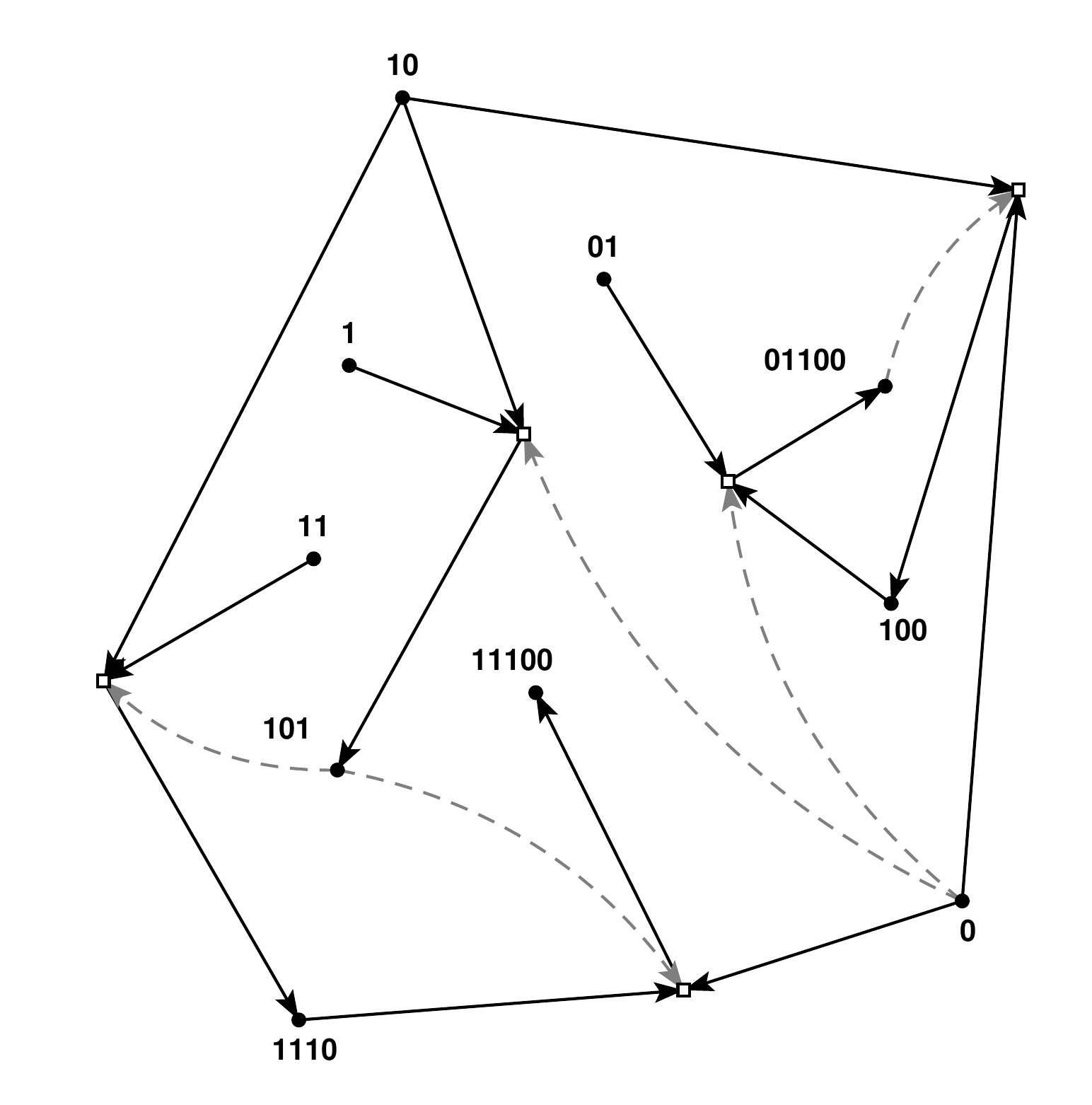}
\caption{Example a RAF set generated in an instance of a binary polymer model, where polymers consist of two monomer species '0' and '1'. The food set is $F = \{0, 1, 01, 10, 11\}$. Molecule types are represented by black dots and reactions by white boxes. Solid arrows indicate reactants and products coming in and out of a reaction, and dashed arrows indicate catalysis. Figure from \cite{HSK2012}.} \label{fig:RAF}
\end{figure}

This formal definition of a RAF set is meant to capture the notion of ``catalytic closure'' of a self-sustaining set of molecules. It is assumed that the the food set F contains molecules freely available in the environment. An example of a RAF set is shown in Fig. \ref{fig:RAF}. RAF theory builds on earlier proposals of autocatalytic sets, including Kauffman's \cite{Kauffman1986}, which envisioned that an autocatalytic network would ``crystallize'' if a sufficient number of catalysts were present in a chemical reaction network \cite{Kauffman1986, Kauffman1993}. In the original Kauffman model, the hypothesis was that autocatalytic sets would be inevitable once a sufficient diversity of molecular building blocks were present. The model considered catalyzed formation and degradation of polymers, composed of $B$ different monomer species.  While the number of different polymer types increases exponentially with length $L$ as $B^L$, the number of reactions necessary to generate a polymer of length $L$ increases even more rapidly (there being many more routes to forming longer polymers than shorter ones). The ratio of reactions to polymers therefore grows linearly as $\sim L-2$. Assuming that every polymer is a catalyst for a reaction with some fixed probability $P$, than as the diversity of molecular species increases it should be the case that eventually the likelihood of an autocatalytic set arising should approach $\sim 1$ in a graph theoretic phase transition. Farmer {\it et al.} estimated the critical probability for this phase transition to occur to be \cite{FKP1986}:
\begin{eqnarray}
P_c \approx B^{-2L}.
\end{eqnarray}
However, this model turns out over-estimate the likelihood of forming an autocatalytic set, and as pointed out by Lifson, would require an exponentially increasing level of catalysis with system size \cite{Lifson1997}. It was later shown within the context of RAF theory that only a quadratic \cite{Steel2000} or even a linear \cite{HS2004,MS2005} growth rate in the level of catalysis is in fact necessary to find autocatalytic sets, which may not be unreasonable, resulting in an increased interest in realizing theoretical and experimental models of autocatalytic sets in recent years. 

\subsubsection{Evolvability of Autocatalytic Sets.}
The above described theoretical approaches have primarily focused strictly on the graph-theoretic aspects of autocatalytic sets, basically whether there exist large connected components within directed reaction networks. More recent theoretical work is moving toward understanding the kinetic properties of catalytic networks and their evolvability.  Graph-theoretic approaches have demonstrated that many autocatalytic networks are hierarchically organized, with ``sub-RAFs'' occurring within larger RAF sets. This suggests potential evolutionary pathways whereby one RAF set could transition to another \cite{HSK2012}. A very simple kinetic example was demonstrated in \cite{HS2014}, showing that irreducible RAF sets (irrRAFS), defined as a RAF that cannot be reduced without loosing the RAF property (contains no subRAFs), can spontaneously form, and be outcompeted by subsequent irrRAFs that emerge later (thus suggesting evolutionary progression). The concept of an irrRAF is closely related to that of a ``viable core'', introduced in Vasas {\it et al.} as sub-networks that form units of heredity (as analogs of ``genes'') in the evolution of autocatalytic sets \cite{evolbeforegenes}. In  Vasas {\it et al.}, evolution of compartmentalized autocatalytic sets was demonstrated via computational models. However only a small fraction (0.01\%) showed persistent increases in non-food set mass over the course of the dynamics of the simulation experiments. These were associated with the presence of viable cores and thus heredity and evolution. It  should be noted that the evolutionary potential of such systems is limited by the number of attractors in the chemical reaction space, and therefore it is unlikely that autocatalytic sets alone could support continual open-ended evolution, although this remains to be confirmed (and could require explicit environmental feedback). In a separate study, Filisetti {\it et al.} also found the spontaneous formation of autocatalytic sets in a stochastic kinetic model to be a rare occurrence \cite{Filisetti2011}. They note a structural fragility of autocatalytic sets -- rare reactions can prevent catalytic closure -- and cite this fragility as a potential reason autocatalytic sets have been so difficult to detect over background chemistry in wet-lab experiments. 

It is important to note that this same issue of continued evolvability also plagues genetics-first models, although it is much less widely recognized due to the tendency to impose biological trends on chemical systems. It remains to be demonstrated whether a network of genetic polymers will similarly ultimately converge to a set of attractors, defined by their relative fitness, or if open-ended evolution is indeed possible, and if so under what conditions. Thus far, no examples of open-ended evolutionary systems are known outside of natural examples from biology and technology. The problem of continued growth in complexity via open-ended processes is therefore dubbed a ``millenial prize problem'' in the artificial life community \cite{OEE}. A resolution will undoubtably have important implications for understanding of the origins of life \cite{RMPM2004}, in particular by inform whether simple chemical networks can increase in complexity in an open-ended way or if additional structure is necessary. 

\subsubsection{Experimental Models.}
In addition to theoretical advances, in recent years autocatalytic sets have been demonstrated in the laboratory in a variety of chemical systems. The original model of Kauffman considered the self-organization autocatalytic sets of proteins, and indeed a autocatalytic network of small peptides has now been confirmed in the laboratory \cite{AJYG2004}. There have also been experimental demonstrations of autocatalytic sets of nucleotide-based polymers, including two-member sets \cite{SvK1994,LJ2009}. Recently a sixteen member set of RNA ribozymes has been experimentally demonstrated to spontaneously self-organize \cite{NVetal}, and was been formally demonstrated to be network autocatalytic \cite{HS2013}. 

\subsubsection{Limitations of Autocatalytic Networks}
While promising, there is much work to be done to demonstrate that cooperative networks of molecules could have been the first evolutionary systems on the pathway to life. We need to understand better both the circumstances under which cooperative networks can spontaneously form from random sets of catalyzed reactions (rather than engineered sets as has been done in the lab or simple computational experiments) and the conditions under which such networks can be said to evolve. Defining heredity, for example, remains a challenge \cite{heredity}. Focusing on identifying the parameters governing the self-organization and dynamics of catalytic networks should shed light on these questions. In Nghe {\it et al.}, we identified six key parameters in need of further investigation: the connectively kinetics of catalysts, the concept of viable cores (or irrRAFs), information control and transfer, network topology, resource distribution patterns, and the role of compartments, see \cite{Ngheetal2015}. In particular, a key challenge in identifying mechanisms of heredity is to better understand how dynamic networks store and process information \cite{Kimetal, Walkeretal}. An example, attempting to unify the concept of catalytic networks with the emergence of genetic systems is provided by Kaneko and collaborators \cite{Kaneko2002, Kaneko2010}. In their model, slow reactions produce ``minority molecules'' that in turn control the reproduction rate of the entire network, acting like a genetic core. Thus a few molecules act to store information about reproduction of the entire system. More work remains to be done to understand the dynamic emergence of separation of information storage and processing and its role in the evolvability of networks and the role of controllability \cite{Kimetal}. 

\subsection{Testing Alternative Hypotheses}

It would be impossible to review all of the proposed models for life's origins herein. However, it should be clear from the forgoing discussion that each ``theory'' for the origin of life must adopt a working definition for life. For approaches exploring the historical sequence of events leading to the emergence of life on Earth, it is important to keep in mind that these approaches provide proof of concept models for abiotic synthesis and are not necessarily reconstructions of events as they actually happened on the primitive Earth. At present, we do not know yet which definition--theories (or combinations thereof) may or may not turn out to be universal. For example, while it is true that life on Earth relies on genetic heredity mediated by linear informational polymers, it is not {\it a priori} obvious that this is necessary a universal feature of life. Likewise for models assuming autocatalysis, compartmentalization, {\it etc.} occurred first. 

\begin{figure}
\centering
\includegraphics[width=5in]{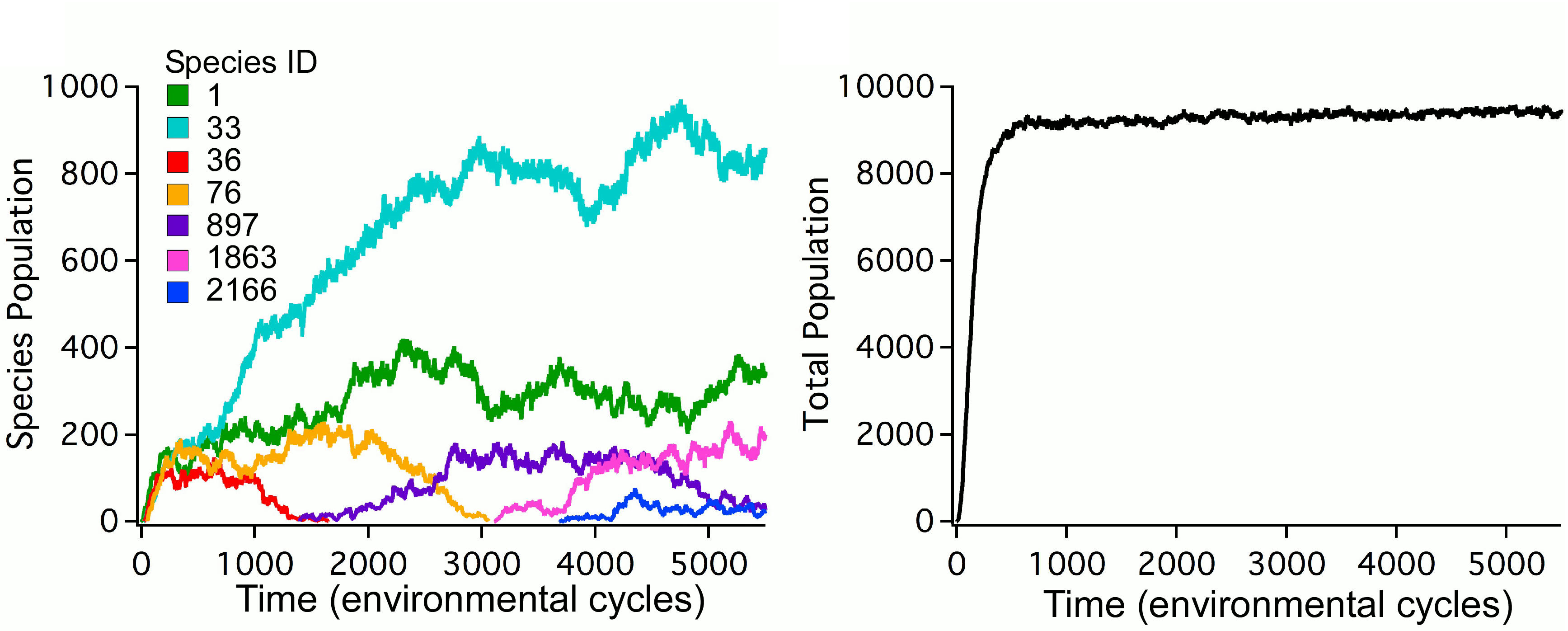}\\
\includegraphics[width=5in]{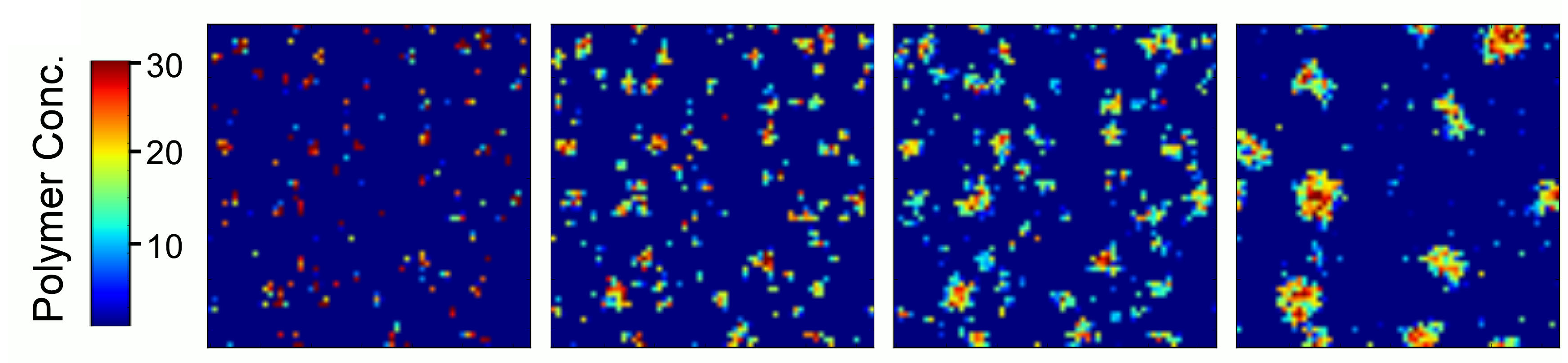}
\caption{Top: Illustration of the concept of dynamic kinetic stability, where the total population of replicators is constant (left) but there is continual turnover of members in the population (right). Bottom: Spatial patterns emerging due to feedback between replicator populations and available resources taken at time points from the dynamics shown in the top panel. Figure adopted from \cite{WGH2012}.} \label{fig:DKS}
\end{figure}

One cannot  formulate a question related to life's origins without assuming something about life, so the issue of defining life ultimately cannot be avoided (although one need not acknowledge it explicitly or make it a focus of research). The question is then, how can the question of the origin of life be asked constructively? For one, there is a need to put the alternative hypotheses for how life emerged on equal footing and test their relative viability. So far, they models for the origins of life have been assessed as independent hypotheses. In particular, computational models could provide a guide for testing competing hypotheses {\it in silico}, since laboratory experiments are often more difficult and more expensive. For example, if replicators and autocatalytic sets occupy the same abstract chemical space in a computational experiment, which is more likely to arise spontaneously - and under what assumptions? Could this then be tested in the lab? It may likely turn out that these hypotheses are not wholly independent -- for example, catalytic sets and replicators both rely on the concept of {\it kinetic} (rather than thermodynamic) stability, which is most often discussed only in the context of replicator models \cite{Pross2004, PPS2013}, see the top panel of Fig. \ref{fig:DKS}. {\it Dynamic kinetic stability} leads to persistence even for short lived entities due to fast kinetics of reproduction. There are a number of models in recent years that begin to blur the boundary between 'genetics' and 'metabolism', for example in metabolic-replicator models as in \cite{WGH2012, KS2011, WP2010}. In metabolic-replicator models, populations of replicators have activity that changes their environmental context. An example of spatial patterning formation arising due resource limitation in replicator populations is shown in bottom panel of Fig. \ref{fig:DKS}. A challenge for future research is to under what circumstances the apparently diverse facets of alternative hypotheses for life's origins might be synthesized. It may be that our current definitions are not truly orthogonal. 

\section{Life from known physics} \label{Sec:phys}

Ultimately, it is unlikely that we will resolve the issue of the origin of life in the absence of understanding what life is. Since Schr\"odinger's seminal set of lectures titled {\it``What is Life?''} published in 1944 \cite{S1944}, there has been a widespread expectation among physicists that an explanation for life will be uncovered with an adequate theory of non-equilibrium physics. Schr\"odinger himself eluded to the role of extensions of thermodynamic theory in addressing the question posed in the title of his lecture series in his coining of the term ``negentropy'' to describe how, seemingly paradoxically,  biological systems `defy' the 2nd law of thermodynamics to increase order locally (sometimes referred to as Schr\"odinger's paradox). Of course, when considered as open systems there is no paradox associated with violating the 2nd law, but nonetheless the question remains - {\it what is life?} After reviewing what was known about living systems at the time, Schr\"odinger conceded  
\begin{quotation}
{\it `` \ldots living matter, while not eluding the ``laws of physics'' as established up to date, is likely to include ``other laws of physics'' hitherto unknown, which however, once they have been revealed, will form just as integral a part of science as the former.''}
\end{quotation}
It has now been over 70 years since the publication of {\it``What is Life?''} and Schr\"odinger's appeal to ``other laws of physics''  remains to be resolved -- we do not know if life can be adequately explained by known physics, or whether additional principles are necessary. In this section I review how our current approaches to understanding physical systems, particularly in statistical physics, can inform investigations into the origin of life and shed light on our understanding of life as a physical process. In the next section, I depart from the view that these approaches alone will ultimately be sufficient to solve the origin of life by providing motivation that new principles may indeed be necessary. 

\subsection{Life as metabolism} \label{sec:met}
A common feature of the autocatalytic networks described in Section \ref{sec:auto} is that the chemistry of interest often consists of linear, combinatorial polymers such as peptides or nucleic acids. However, the math is sufficiently general that this need not necessarily be the case. For example, autocatalytic sets have been found within the metabolic network of {\it Escherichia coli} \cite{SHSM2015}. Autocatalytic networks can in principle apply to any collectively reproducing set of molecules (or other collectively reproducing entities such as economic firms \cite{Hordijk2013}). There has been much discussion in the origin of life literature about what molecules might have contributed to the first metabolisms and whether sets of molecules constituting a metabolism could collectively reproduce and jump-start an evolutionary process. Thus, expanding the idea of collective reproduction makes contact with another class of origin of life theories associated with the early emergence of energy transduction pathways.  However, here the focus is not on the kinetic organization {\it per se}, but instead the organization of energy flows \cite{MS2007} (in reality both are tightly coupled \cite{CSM2010}). Under this view, life emerged in order to release thermodynamic stresses on the primitive Earth. This perspective defines life as a direct outcome of planetary geochemistry \cite{ShockBoyd2015,SmithMorowitz2004}. 

\subsubsection{Biochemistry as the ``chemistry that Earth allows''}

Life requires free energy. This had led many researchers interested in life as a mechanism for energy transduction to seek out potential environments on early Earth that would provide ample free energy. Popular among this set of hypotheses is that life first emerged in a hydrothermal vent system \cite{Martin2008}, perhaps not too unlike the black smokers found in Lost City \cite{Hoffman1981}. That is, life should have emerged on Earth where there were opportunities for catalysis to expedite the release of chemical energy, for example in water--rock--organic systems \cite{ShockBoyd2015}. An important question is: what biomolecules could be produced under such conditions, and could is chemical synthesis of organic molecules be thermodynamically favored? 

Much research in this direction has focused on the ``iron-sulfur'' world hypothesis as first proposed by W\"achtersh\"auser \cite{Wach1988}, wherein life is proposed to have first emerged in a hydrothermal vent at high pressure and high temperature. Under these conditions transition metals (such as iron and nickel) can act as catalysts for synthesis of small organic compounds from inorganic gases. This surface based, carbon-fixation metabolism could become network autocatalytic through the formation of a metabolic cycle in the form of an ancestral sulfur-dependent version of the reductive citric acid cycle (reverse TCA) \cite{Wach1990}. Support for this view comes from empirical evidence that metabolism has a common core in the form of reverse TCA \cite{MKYG2000}. This universality may be a solution imposed on life within energetically structured environment of early Earth \cite{SmithMorowitz2004}. Once a simple autocatalytic metabolism was established, it is proposed that chemical synthesis reactions could then produce more complex organic compounds, bifurcating down energy-releasing pathways to produce molecules of increasing complexity (see {\it e.g.} \cite{BS2012}). Genetic systems would have emerged later, as a product of these kinds of synthesis reactions. Many elementary steps in the iron-sulfur world hypothesis have been experimentally confirmed \cite{MR2003, HW1998}. Additionally, amino acids and short peptides have been demonstrated to be thermodynamically favorable under conditions of high T and P \cite{ShockAmend}, suggestive that the biomolecules of life might be readily produced in certain geochemical contexts. This has led to the idea that ``biochemistry is what the earth allows'', in other words, that life emerges as a planetary response to trapped energy as the Earth cooled (see \cite{ShockBoyd2015} for discussion). In this picture, the biosphere emerged as a direct outgrowth of the geosphere \cite{SM2016}, driven by energy flows that generated structure at multiple hierarchical scales \cite{MS2007}.

\subsection{Life from thermodynamics} \label{sec:SMrep}
In the more abstract, one can consider how thermodynamics might drive the emergence of life in simple models. Although similarly focused on thermodynamics, this shifts emphasis from regarding life as a product of geochemistry to focusing on life as the emergence of reproducible organized structures, as occurs via self-replication, and its thermodynamic consequences. 
The idea that life might exist as a dissipative structure became popular with the work of Prigogine and Nicolis (see {\it e.g.} \cite{PN1971}) and has been applied to origin of life models  \cite{Virgo}. A core underlying idea is that a principle based on maximizing entropy production could drive the formation and subsequent evolution of living systems \cite{Whitfield}.  Recently this kind of approach was adopted by England to study dissipation in the process of self-replication \cite{England2013}. Self-replication is a statistically irreversible process: a single cell can replicate to produce two daughter cells, but we do not observe the reverse situation in which two `daughter' cell spontaneously convert into one. In equilibrium thermodynamics, irreversibility is accompanied by an increase in entropy. It follows that self-replication (as a far-from equilibrium and irreversible process) might also produce an increase in entropy. 

Here I follow the example of England in his rough sketch of the entropy production for a self-replicator coupled to a heat bath \cite{England2013} to illustrate the approach, which leverages recent advances in non-equilibrium thermodynamics through the development of fluctuation theorems \cite{fluctuation, Crooks1999}. The most general form is the Evan Searles fluctuation theorem (FT), which relates the probability of a trajectory and its reverse to entropy production. For driven systems this is nonzero:
\begin{eqnarray} \label{CFT}
\frac{{\cal P}(a \rightarrow b)}{{\cal P}(b \rightarrow a)} = e^{\beta \Delta Q (a \rightarrow b)}
\end{eqnarray}
where $\beta = 1/T$ is the inverse temperature of the heat bath, ${\cal P}(a \rightarrow b)$ is the probability of the system transitioning from microstate $a$ to microstate $b$ in a forward direction, ${\cal P}(b \rightarrow a)$ is the probability of the reverse trajectory from $b$ to $a$, and $\Delta Q (a \rightarrow b)$ is the heat released to the bath over the forward path from $a$ to $b$. The essential implication of Eq. \ref{CFT} is that the more irreversible the process ({\it i.e.}, the larger the inequality ${\cal P}(a \rightarrow b) \gg {\cal P}(b \rightarrow a)$), the more heat is dissipated into the surrounding universe. 

Fluctuation theorems are typically applied to microscopic processes. Life, by contrast, is a macroscopic phenomenon. Eq. \ref{CFT} can however be implemented to understand the irreversibility of {\it macroscopic} observables of the system. These can be any coarse-grained variable describing the system. Here we consider two arbitrary macrostates ${\cal A}$ and ${\cal B}$. Following England, we assume that if we observe the system in macrostate ${\cal A}$, we can associate a conditional probability $p(a|{\cal A})$ that the system was a particular microstate $a$. Likewise if we instead observe the system in state ${\cal B}$, we associate the conditional probability  $p(b|{\cal B})$ that the system was in microstate $b$. For example, macrostates of interest might be a state ${\cal A}$ consisting of one living cell, and macrostate ${\cal B}$ consisting of two living cells. Assuming a macroscopic description of the irreversibility of spontaneously transitioning from ${\cal A}$ to ${\cal B}$ \cite{England2013}:
\begin{eqnarray}
\frac{{\cal P} ({\cal B} \rightarrow {\cal A})}{{\cal P} ({\cal A} \rightarrow {\cal B})} = \left\langle \frac{\langle e^{- \beta \Delta Q_{ab}}\rangle _{a \rightarrow b}}{e^{\ln \left[ \frac{p(a|{\cal A})}{p(b|{\cal B})} \right]}} \right\rangle _{{\cal A} \rightarrow {\cal B}} 
\end{eqnarray}
where $\langle \ldots \rangle _{{\cal A} \rightarrow {\cal B}}$ denotes an average over all paths from some microstate $a$ in the initial ensemble to some microstate $b$ in the final ensemble. This is somewhat problematic as the procedure for determining what microstates to include in the ensemble on the left-hand side of the equation is restrictive: it is not {\it a priori} obvious that this is a natural partitioning emerging from the system's dynamics rather than one we've imposed on the system ({\it i.e.} identifying ${\cal A}$ and ${\cal B}$ as one cell and two cells is somewhat arbitrary).  Rearranging yields a generalization of the 2nd law of thermodynamics that applies to the irreversibility of macroscopic processes (see \cite{England2013} for details):
\begin{eqnarray} \label{secondlaw}
\beta \langle \Delta Q \rangle_{{\cal A} \rightarrow {\cal B}} + \ln \left[ \frac{{\cal P} ({\cal B} \rightarrow {\cal A})}{{\cal P} ({\cal A} \rightarrow {\cal B})} \right] + \Delta S_{\rm int} \geq 0
\end{eqnarray}
where $\Delta S_{\rm int}$ is the entropy change of the system given by $\Delta S_{\rm int} \equiv S_{\cal B} - S_{\cal A}$ with $S = - \sum_i p_i \ln p_i$. In the event that ${\cal P} ({\cal A} \rightarrow {\cal B}) = {\cal P} ({\cal B} \rightarrow {\cal A}) = 1$, such that both macrostates share an identical set of microstates, the above reduces to the 2nd law in a more familiar form: the total entropy change of the system $\Delta S_{\rm int}$ and of the heat bath $\beta \langle \Delta Q \rangle_{{\cal A} \rightarrow {\cal B}}$ must be positive. Macroscopic irreversibility sets a stricter bound, with more irreversible macroscopic processes resulting in larger minimum entropy production. 

It should be stressed that nothing about Eq. \ref{secondlaw} is specific to life or replication, but applies to {\it any} irreversible macroscopic transformation. It therefore does not hold any insights that are specific to life's emergence that don't apply to dissipative processes more generally (it is an open question whether there are such processes). Eq. \ref{secondlaw} can be however be directly applied to self-replication by considering a population of simple self-replicators at inverse temperature $\beta = 1/T$ whose replication and decay rates are given by $g$ and $\delta$, respectively. In an infinitesimal time $dt$ the probability that a replicator in the population reproduces ${\cal P} ({\cal A} \rightarrow {\cal B})$ is then $g dt$ and that of a replicator decaying ${\cal P} ({\cal B} \rightarrow {\cal A})$ is $\delta dt$. Plugging these transition probabilities into Eq. \ref{secondlaw} and rearranging yields:
\begin{eqnarray} \label{eqn:rep}
g_{\rm max} - \delta = \delta (e^{\beta \Delta q + \Delta S_{int}} - 1)
\end{eqnarray}
where it is assumed that $g > \delta$, thus lower bounding the total entropy produced via self-replication. This framework is sufficiently general to apply to self-replicating molecules and catalytic networks, which could likewise be cast in terms of forward and backward rates for synthesis. All else being equal, the thermodynamic benefits of self-replication quantified by Eq. \ref{eqn:rep} seem to favor the simplest replicators ({\it i.e.} the shortest replicators which can replicate and degrade the fastest and therefore maximize entropy production). However, this misses a critical point about information and its role in selection of replicators -- all else is not equal. Physical systems encoding the information necessary to replicate fast will do so at an exponential rate \cite{Carothers}, whereas sequences of similar length that contain no fitness-relevant information will die. That information and selection matter to life has been one of the most challenging aspects of understanding life as a physical process, and nonequilibrium approaches have yet to address this issue -- even if we could identify natural or ``intrinsic'' macrostates. The forgoing demonstrates that selection for systems that dissipate energy at a fast rate will yield simple replicators. Dissipation is a consequence of selection of information, not a driver of it. Co-polymerization provides one explicit example where dissipation is closely related to information \cite{AG2008}. It seems likely that in the absence of appealing to informational principles, discussions of dissipation and entropy-production alone cannot explain the origins of life (hence Schr\"odinger's original appeal to ``other laws''). 

\subsection{Life as a critical phenomenon.}

Another set of ideas carried over from physics to the study of living processes are those associated with criticality and phase transitions.  In particular, criticality has been proposed in a variety of contexts as essential to the adaptability of living matter, where living systems are often argued to be poised at the critical point between order and disorder (colloquially this is sometimes phrased as ``poised at the edge of chaos'') \cite{Adami95, KJ1991, Lewin1999, Langton1990, Packard1998}.  It is an open question how criticality in biological systems maps to our understanding of similar concepts in physics. The theory of physical phase transitions encompasses a broad class of critical behavior, ranging from the transition from liquid to solid phases to the phase transition  associated with spontaneous symmetry breaking to form magnetized domains in ferromagnets. In biology, criticality is most often associated with the properties of networks and information flows within those networks (see {\it e.g.} \cite{Daniels}). Criticality is also directly applicable to the problem of the origins of life, where the mathematics used to describe symmetry breaking processes and phase transitions provides useful mathematical tools. 

\subsubsection{Homochirality.} \label{Sec:Chiral}

One area where the tools of physics find direct application is in explaining the emergence of homochirality, which readily lends itself to description in terms of spontaneous symmetry breaking.  Many of the universal biomolecules listed in Table \ref{Table:univ} come in two chiral forms much like your left and right hand are mirror images of each other (``chiral'' is derived from the Greek word for hand). Yet, all known life is homochiral \cite{homochiral}: life uses primarily left-handed amino acids and right-handed sugars in DNA and RNA. This is in contrast to what is found abiotically: both left- and right-handed chiral forms (enantiomers) are formed under prebiotic conditions (although meteorites often have small asymmetries favoring some left-handed amino acids, see {\it e.g.} \cite{CP1997}). Some symmetry breaking process(es) must have occurred during the emergence or early evolution of life to give rise to the asymmetry of the biosphere observed today.

While there exist many models describing how homochirality may have emerged in prebiotic systems, the qualitative features of the majority of these models are similar: chiral symmetry breaking occurs due to the introduction of instabilities to the symmetric state (containing both chiral forms) that lead to spontaneous symmetry breaking in physical systems \cite{Frank1953, Kondepudi1987, Hochberg2010}. The spatiotemporal dynamics of a chiral reaction network can be equated to a two-phase system undergoing a symmetry-breaking phase transition, where the order parameter is the net chiral asymmetry ($A$) \cite{BM2004,G2007}. Defining $L$ and $R$ as the sums of all left and right-€"handed chiral sub-units, respectively, the net chirality may be defined as
\begin{eqnarray}
A = \frac{L-R}{L+R}
\end{eqnarray}
The net chirality is symmetric $A = 0$ for $L=R$ (the racemic state), and asymmetric $A \neq 0$ in the non-racemic states (an excess of $L$ or $R$). 

In general, reaction networks are nonlinear dynamical systems with behavior controlled by model-dependent parameters. Parameters relevant to generating chiral asymmetry include fidelity of enzymatic reactions \cite{Sandars2003, SH2004}, ratios of reaction rates \cite{GW2008}, stereoselectivity \cite{Plasson2004}, total system mass \cite{GW2009} and even stochastic noise \cite{Jafarpour, SHS2008}. Here, we consider an explicit example where chiral symmetry breaking is controlled by reaction fidelity to illustrate general mechanisms, utilizing the set of heterochiral polymerization reactions as proposed by Sandars \cite{Sandars2003}:
\begin{eqnarray}\label{eqn:sandars}
L_n + L_1 \stackrel{2 k_S}\longrightarrow L_{n+1}\\
L_n + R_1 \stackrel{2 k_I}\longrightarrow L_{n} R_1 \label{eq:kI} \\
L_nR_1 + L_1 \stackrel{k_S}\longrightarrow L_{n+1} R_1\\
L_nR_1 + R_1 \stackrel{k_I}\longrightarrow R_1 L_{n} R_1
\end{eqnarray}
where $k_S (k_I)$ are the reaction rates for adding sub-units (monomers) of the same (opposite) chirality to a growing polymer, and $L_n$ and $R_n$ denote left- and right- handed polymers $n$. A mirror-image set of reactions hold for $L \leftrightarrow R$. An essential feature of this set of reactions is the feedback inhibition associated with attaching a monomer of the wrong-handedness to a growing polymer. Thus, for example, in Eq. \ref{eq:kI} the attachment of $R_1$ to the end of a growing $L$-polymer will terminate growth at that end of the polymer, a process termed cross-inhibition. Such enantiomeric cross-inhibition is an essential aspect of producing homochirality in the original Frank model \cite{Frank1953}, from which the majority of models for chiral symmetry breaking derive. Frank showed that inhibition, when coupled with autocatalytic feedback, provides a sufficiency condition for producing homochirality from a nearly racemic initial condition (see {\it e.g.} \cite{Jafarpour,SHS2008} for two recent examples that do not need to include inhibitory feedback and instead rely on noise). The above set of reactions are therefore complemented by autocatalytic creation of monomers by the following two reactions:
\begin{eqnarray}
S \stackrel{k_C [L_N]} \longrightarrow L_1\\
S \stackrel{k_C [R_N]} \longrightarrow R_1
\end{eqnarray}
structured such that long homochiral polymers catalyze production of monomers of the same chiral species. In this example, the longest polymers formed (length $N$) are the only sequences that act as catalysts. 

\begin{figure}
\centering
\includegraphics[width=3in]{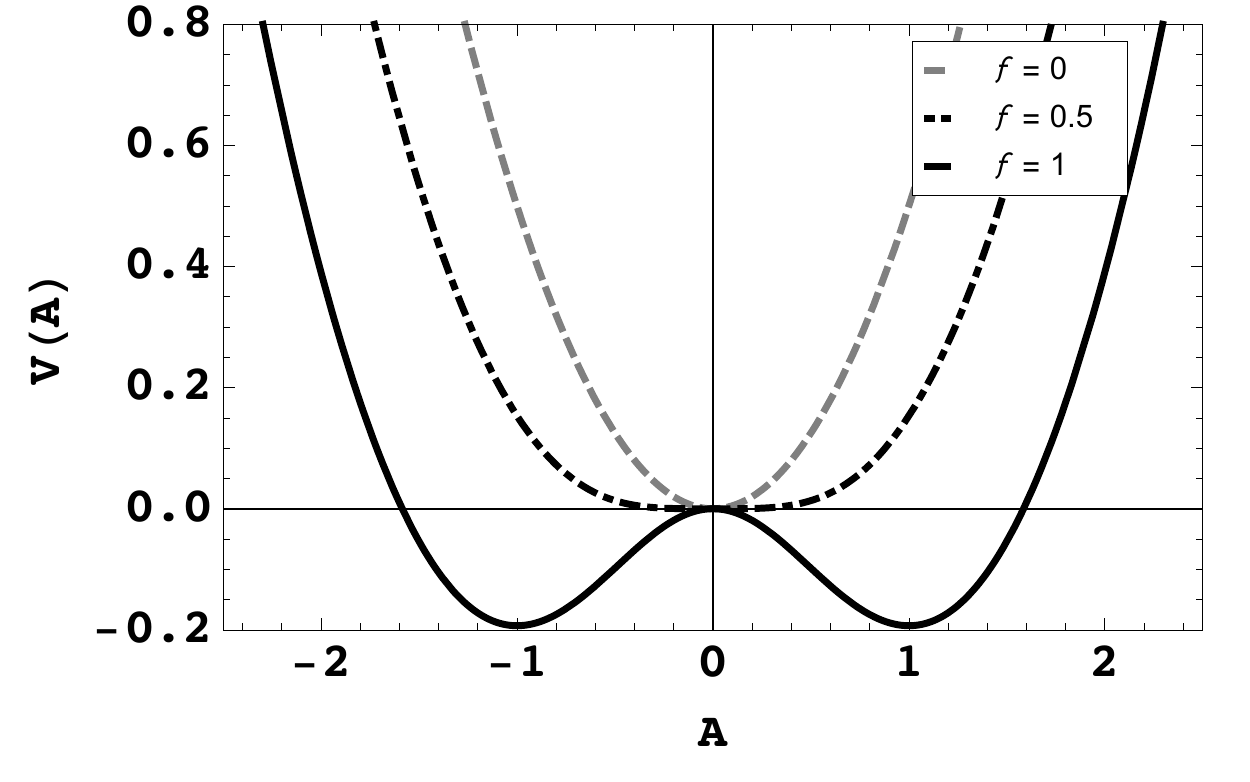}
\includegraphics[width=2in]{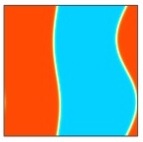}
\caption{Left: Chiral symmetry breaking arising due to varying values of the control parameter $f$, governing the fidelity of enzymatic reactions. Right: An example of domain wall formation in the spatiotemporal dynamics of the reaction network for $f=1$. Orange (L-phase), Blue (D-phase), White (racemic)} \label{fig:pot}
\end{figure}

The Sandars model can be truncated to $N=2$ and still maintain all essential features of the dynamics leading to homochiralization \cite{BM2004}, allowing one to model the reaction network utilizing tools of mean-field theory to characterize spontaneous chiral symmetry breaking \cite{G2007}. Introducing a few assumptions (such as the rate of change of dimers is slow compared to that of monomers), it can be shown that the above set of reaction equations reduces to:
\begin{eqnarray}
\lambda_0^{-1} \frac{d {\cal S}}{dt} = 1 - {\cal S}^2 \label{eqn:S}\\
\lambda_0^{-1} \frac{d {\cal A}}{dt} = 2f \frac{{\cal SA}}{{\cal S}^2 + {\cal A}^2} - {\cal S} {\cal A} \label{eqn:A}
\end{eqnarray}
where $\lambda_0 \equiv (2 k_S Q)^{1/2}$, with dimension of inverse time. Here ${\cal S} \equiv X + Y$ and ${\cal A} \equiv X - Y$ are symmetric and asymmetric variables describing the total mass and net asymmetry, where $X \equiv [L_1] (2 k_S / Q) ^{1/2}$ and $Y \equiv [R_1] (2 k_S / Q) ^{1/2}$ (see {\it e.g.} \cite{GW2008} for discussion).  By Eq. \ref{eqn:S}, ${\cal S} = 1$ is a fixed point of the dynamics, which the system relaxes to with a characteristic timescale $\lambda_0$. Setting ${\cal S} = 1$ in Eq. \ref{eqn:A}, the equation dictating the net chiral asymmetry, one arrives at the effective potential for ${\cal A}$:
\begin{eqnarray}
V({\cal A} )= \frac{{\cal A}^2}{2} - f \ln \left[ {\cal A}^2 + 1 \right]
\end{eqnarray}
As shown in Fig. \ref{fig:pot}, for $f < 0.5$ this potential has a characteristic double-well shape, with minima at the fixed points ${\cal A } = \pm \sqrt[]{2f - 1}$. By the shape of the potential, it is evident that $f_c = 0.5$ is the critical fidelity for spontaneous symmetry breaking: fidelity $f < 0.5$ will lead local domains to assume left- or right-handed chirality, as shown in the right panel of Fig. \ref{fig:pot}. 

Salam first suggested that there should be a critical temperature, $T_c$, above which any net chirality is destroyed in a prebiotic system. In situations where $f=1$ the system can be restored to a racemic (symmetric) state by coupling it to a heat bath \cite{GT2006}.  There is a direct analogy with a ferromagnetic phase transition (where ferromagnets likewise can take on one of two states -- up or down spin): if heated through the Curie point any net magnetization is erased in a ferromagnet and the system is restored to a symmetric configuration. Here, the net chirality plays the role of the net magnetization. Repeated thermalization can reset the chirality of a system \cite{GTW2008}. Net chiral asymmetry is therefore a candidate that may be among the order parameters of a ``phase transition'' to the living state. What remains to be determined is whether homochirality is a universal feature of life (which has implications for using homochirality as a biosignature, for example in searches for life on Mars). A related question is whether homochirality preceded life or is a byproduct of it \cite{WWH2012}, in other words did this kind of symmetry breaking occur before, after or during the emergence of life? 

\subsubsection{Other examples.}

\begin{figure}
\centering
\includegraphics[width=4.5in]{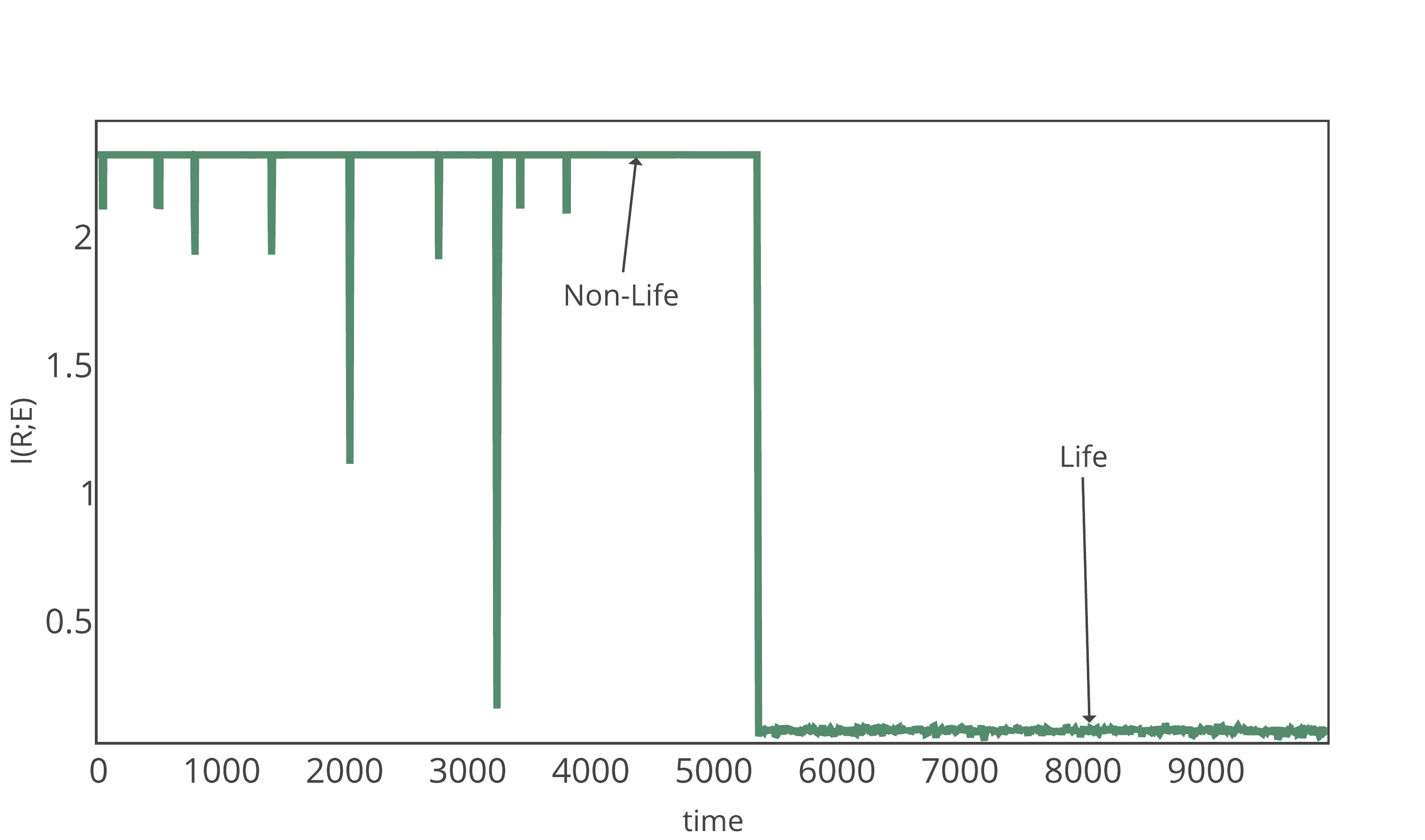}
\caption{Time series of the mutual Information shared between replicators and their environment (free monomers), demonstrating an abrupt phase transition from no replicators (non-life) to dynamics dominated by the selection of replicators (life). Figure adopted from \cite{Mathis2016}.} \label{fig:phase}
\end{figure}

Net chiral asymmetry is just one of potentially many order parameters associated with the emergence of the living state, which remain to be identified. These could be associated with how energy flows organized increasingly `life-like' chemistry as discussed in Section \ref{sec:met}. Some may also be associated with the emergence of life as a ``kinetic state of matter'' \cite{Pross2004, PPS2013}: that is, as far from-equilibrium systems driven by self-replication. In recent years, there has been increased interest in identifying the emergence of replicators from prebiotic systems as a phase transition \cite{Mathis2016, NO2008, MCN2010, WH2009, MK2016}. Nowak and collaborators, for example, have quantified an abrupt `phase transition' from ``pre-life'' to ``life'' associated with the emergence of replicators with high fidelity \cite{NO2008}. In their framework, ``pre-life'' is a generative chemistry capable of producing a diversity of molecules (structured much like the polymerization equations in Eq. \ref{eqn:sandars}).  ``Life'' emerges with polymers that can copy themselves, such that they can replicate and evolve. Here the fidelity of replication serves as the order parameter: if replicators appear that are effective at self-copying the system will select replicators, otherwise it will stay in the ``pre-life'' phase and favor polymerization. A catch-22 with this simple model is that while easy to tune parameters in a model, it is not {\it a priori} obvious what selective pressure could drive high replicative fidelity {\it before} replicators emerge, however high fidelity is necessary to mediate the transition. We reported a similarly abrupt, but spontaneous transition in Mathis {\it et al.}, where the relevant order parameter is associated with the tuning of replicators to their environment \cite{Mathis2016}. In this model, the mutual information between replicators and their environment accurately tracks the progress of the phase transition, as shown in Fig. \ref{fig:phase}.
Exploring highly dynamic recycling chemistries, the system will abruptly transition from polymerization dominated dynamics to selection on replicators. The transition is accurately captured by tracking the mutual information between replicators and their environment. In all of these models, it remains an open question whether replicators might be identified as a separate ``kinetic state of matter'' and whether the dynamics described in a variety of different models represent a true phase transition. 

\section{Life from new physics?} \label{Sec:new_phys}

We have now seen a number of explicit examples modeling the emergence of life, and how the traditional tools of physics can come to bear on the problem. However, even with so many distinct hypotheses and approaches to the problem, we are far from a resolution to our question -- {\it ``How is it that life can emerge from non-living matter?''}. Most work on fundamental properties of life focuses on the concept of information \cite{Bialek2012, DW2016} and this may also be critical for quantitative theories of life's origins \cite{WD2013}. There are some hints toward a potentially deep connection between information and thermodynamics due to the mathematical relationship between Shannon and Boltzmann entropies \cite{Jaynes1957}. Substantial work over the last decade has attempted to make this connection explicit, see {\it e.g.},  \cite{Parrando2015} for a recent review. Schr\"odinger was aware of this link in his deliberations on biology, coining ``negentropy'' to describe life's ability to seemingly violate the 2nd law of thermodynamics. Yet, as mentioned in Section \ref{Sec:phys}, he still felt that something was missing and ultimately resorted to posing that ``other laws'' might be necessary \cite{S1944}. In what remains, I take a more forward-looking approach and speculate on what might be necessary to finally resolve the problem of the origin of life, focusing on the promise of the development of information-based theories.


\subsection{Life and information}

One problematic aspect unique to life (and its artifacts) is the apparent trend of open-ended evolution of the biosphere over its $>3.5$ billion year history. This could be an artifact of post-selection (if the biosphere were not complex we certainly wouldn't be here to observe it, see discussion in Section \ref{sec:anthropic} and also \cite{LDM2013}). Post-selection, while a valid explanation, is unsatisfactory as it doesn't  explain how we came to be. During the emergence of life, a process driving increasing complexity with time was necessary \cite{CroninWalker2016}. Darwinian evolution does not satisfactorily fill this void if taken alone ({\it e.g.} in the absence of feedback between environment and nascent life, see Section \ref{sec:gen_lim}), as it can lead to both simpler or more complex systems depending on the context of selection. In fact, a challenge with the majority of replication-based scenarios for the origin of life is that they either stall-out at a stage of relatively low complexity, or evolve toward states of lower complexity with time ({\it e.g.} as demonstrated by Speigelman's monster, see Section \ref{sec:gen_lim}), specifically because information-rich environments are not included in the discussion \cite{AdamiLabar}.

Belief that progress in biology will come from shifting the conceptual basis from matter and energy flows to the abstract realm of information is becoming increasingly recognized \cite{Nurse2008, Bialek2012, Farnsworth2013, Mayfield, AdamiLabar}. A prominent role for informational concepts exists across all levels of biological organization from cells to societies. In neuroscience we have the example of experiencing a ``train of thought'' in which one thought or sensation from sensory input leads to a further thought in what is felt by a conscious entity as a cause-effect relationship. In the social sciences, cultural and political context is often discussed as being causal to individual actions \cite{DeDeo}. A frequently discussed example in biology is the `information hierarchy' \cite{Flack}. Life on Earth is characterized by nested hierarchies, with new `levels' emerging through major transitions. Major transitions in evolution include the origin of eukaryotes, multicellularity, and eusocial and linguistic societies \cite{SMS1995}. It has been hypothesized that re-organization of information, including new modes of storage and processing, drive these transitions \cite{SMS1995} where the boundary of ``individual'' too may change during the transition \cite{Flack2013}. Traditional evolutionary perspectives have thus far not provided satisfactory explanation for the hierarchy of life (see {\it e.g.} discussion on levels of selection \cite{Okasha}).  If ``laws of life'' exist, we might expect that each emergent 'level' represents a new example of life and should include universally shared properties. That is, we can adopt a view where life is {\it not} a level-specific phenomenon (as one must assume if life is defined by its chemistry).  The challenge is to find universal principles that might equally well describe {\it any} level of organization in the biosphere (and ones yet to emerge, such as speculated transitions in social and technological systems that humanity is currently witnessing, or may one day soon witness). The  unifying conceptual basis across these transitions is {\it information}, and it may in fact be that transitions in information flows (perhaps with the emergence of new constructors or programs, see below) are precisely what drives such jumps in complexity \cite{Walker2012}. 

Focusing on information moves the narrative away from a chemistry-dominated description of life, and may provide our best shot at uncovering universal ``laws of life'' that work not just for biological systems with known chemistry but also for putative artificial and alien life. As far as we know, life requires chemistry, but the properties of the living state emerge from the {\it dynamical} properties of that chemistry, including the temporal and spatial organization of chemical networks and the resultant information flows \cite{Nurse2008}. Hallmarks of life, based on an informational perspective are outlined in Table \ref{Table:hallmarks}. Many of these hallmarks may turn out to be different descriptors of the same physical process(es) pointing to a hidden simplicity in the structure of living systems that remains to be fully explicated. We next discuss a few of these hallmarks, focusing on the concept of non-trivial replication (for more extensive discussion see {\it e.g.} \cite{WD2013, DW2016}). 

\begin{table}
\centering
\begin{tabular}{  p{11cm} } 
{\bf Universality in the physics of known life} \\
\hline 
\hline
  global organization \\  \hline 
    information as a causal agency \\  \hline 
      top-down causation \\  \hline
        analog and digital information processing \\  \hline  
         laws and states co-evolve \\  \hline 
         logical structure of a programmable constructor \\  \hline 
          dual hardware and software roles of genetic material \\  \hline 
          non-trivial replication \\  \hline 
          physical separation of instructions (algorithms) from the mechanism that implements them \\  \hline 
  \hline
\end{tabular}
\caption{Some of the informational hallmarks of life. Table adopted from \cite{WD2013}}. \label{Table:hallmarks}
\end{table}

\subsection{Life as (programmable) constructors}




von Neumann was one of the first to consider the active role information must play in living systems \cite{vN1966}. He recognized that {\it copying} information (even with mutation and selection) is not sufficient to generate the complexity of of living systems (which he hoped to emulate in artificial systems), but instead the concept of {\it constructability} must additionally be introduced \cite{Mayfield}. Copying and construction as introduced by von Neumann are two fundamentally different physical processes, although they may ultimately lead to the same effective outcome - the reproduction of information stored in one physical media in another. In the case of copying, the information is replicated from one media to another of the same physical stuff (or nearly so). Examples include crystal growth (copying atomic lattice structure), nucleic acid replication (copying the sequence of nucleobases), or simple physical systems such at that proposed by Penrose and Penrose that can reproduce their own state \cite{Penrose}. Constructors by contrast perform transformations on physical objects, such that one physical media may be transformed into another. An example is catalysis, where the catalyst is the constructor \cite{Deutsch}.  

A constructor so defined does not in of itself constitute a living system, or even a system capable of the relatively simple task of copying. von Neumann therefore devised the concept of a {\it universal constructor} (UC) based on Turing's ideas of universal computers \cite{Turing1936}. A universal computer is a computer that can compute any computable function. A UC by analogy, is a physical system that can construct {\it any} physical object (within a given universality class of objects) when supplied with sufficient resources to do so. In order to specify which physical system to construct, the UC must be supplied with ``€œinstructions''€ that permit the construction of that object from elementary operations (those permissible by the laws of physics). Reproduction of the physical system occurs when those instructions specify how to construct the UC itself (this provides a physical mechanism for self-referential dynamics \cite{WD2013, DW2016, GW2011}, one of the hallmarks in Table \ref{Table:hallmarks}). It is an open problem whether true universal constructors can in principle exist (see Deutsch for discussion on the concept of the ``constructibility of nature'' \cite{Deutsch2011}). Approximations are known to exist -- for example, the logical architecture of the cell has been equated to that of a UC on numerous occasions (see for example \cite{Danchin2009}). Our current state as a technologically advanced civilization is an even better approximation to a UC, as there are certainly many possible transformations that technological civilizations enable in physical reality, which seem impossible in the absence of technology ({\it e.g.} production of the element Technetium, which does not occur naturally but is in ``high'' abundance on Earth, and launching satellites into space, see \cite{Walker2016} for discussion). This concept, that living systems (and their artifacts) mediate transformations that do not violate knows laws of physics, but are at the same time not predicted by them, may be one of the most fundamental features of life. It suggests that an explanation for life is not in explaining the states themselves, but instead the paths \cite{WD2016}. This view is consistent with an emerging emphasis in nonequilibrium thermodynamics on trajectories rather than states ({\it e.g.} see discussion of fluctuation theorems, Section \ref{sec:SMrep}). 

A more general and less strict concept than that of {\it universal} construction, is the idea of {\it programmable constructors}. Programmable constructors do not necessarily operate on a universality class of objects, but through interaction with other physical systems can be ``programmed'' to perform specific physical transformations \cite{Deutsch}. In physics the idea of a ``program'' is itself not well-defined. For purposes of discussion herein, we may consider these as inputs to a particular physical system or device that produce different outputs. The distinction between copying and programmable construction forms the core of distinguishing between trivial and non-trivial replication \cite{WD2013, Walker2014} -- that is between copying and construction, respectively. Life not only copies information but also uses it to construct itself \cite{Marletto} and can utilize information to construct other objects. It should be noted that this could exclude the possibility of life based on a single biopolymer as it is difficult to envision how interactions (``programs'') within such systems could be set-up to perform different transformations for different inputs (as is the case for coded information in the cell, which is a construct of interaction between two classes of biopolymer). Two polymers (or at least two distinct physical media) may be necessary for the physical separation of instructions (programs) from the mechanisms that implement them \cite{RMUM2008} (see hallmarks). 

The forgoing has important implications for how we model the emergence of life \cite{WD2013}. We should be engineering {\it in vitro} and {\it in silico} models more explicitly focused on {\it information} and how it operates in physical systems. In modeling the origin of life, we should not only think of life as ``information that copies itself'', but consider that this implicitly means ``simple machines that can make slightly more complicated machines'' \cite{CroninWalker2016}. It is only non-trivial replicators that {\it process} information in an active sense, enabling the system's dynamics to (in part) be directed by the current informational state (``program'') of the system. This is the key idea behind the philosophical concept of top-down causation \cite{WD2013, DW2016, AEJ2008}. It is often assumed that the behavior of a physical system can be traced back to the behavior of its components (reductionism). This procedure works well in physical systems that permit a separation of scales, such as when we describe the behavior of an ideal gas in terms of averaged properties of its constituent components, that is in terms of macroscopic variables. Living systems represent by contrast are embedded hierarchies, with complex flows of information between scales of organization \cite{Flack} that do not generally permit this layer-by-layer decomposition of causation (information flows from 'higher' to 'lower' levels). That is, life could be regarded as a hierarchy of `constructors', or at least information flows that mediate which transitions occur and when. It is widely recognized that the procedure of coarse-graining (which defines some of the relevant ``informational'' degrees of freedom) plays a foundational role in how biological systems are structured \cite{Flack2013}, by defining the biologically relevant macrostates. However, it is not clear how those macrostates arise, if they are objective or subjective \cite{SM2003}, or whether they are in fact a fundamental aspect of biological organization. The emergence of life can be re-stated as a problem of explaining how (biological) hierarchies emerge (these should be distinguished, for example, from re-normalization group flows or other `hierarchies' in physics, since in biology the individual 'levels' are not self-similar). The mechanisms through which top-down causation, if indeed it is a real and not just apparent property of nature, could operate in biology would most likely be through information (in an as yet unspecified manner) acting as a causal agent (another hallmark in Table \ref{Table:hallmarks}). The idea of information is itself abstract, but it must be the case that each bit of information is instantiated in physical degrees of freedom: ``information is physical!'' in the words of Rolf Landauer \cite{Landauer1992}. Whether fundamental or an epiphenomenon, the causal role of information in biology represents one of the hardest explanatory problems for solving the origins of life \cite{WD2013}.

\section{Conclusion}

Over the last four-hundred years of physics as a scientific discipline, we have made tremendous progress in advancing our understanding of the smallest and largest scales in the universe. However, we have made far less headway at the scales of our everyday experience -- in the realm of the complex and the biological. We understand more about the structure of an atom, something that we do not directly experience, than we do about how complex physical systems such as yourself should arise and be capable of comprehending this page of text in a meaningful way.  Einstein's thoughts on the matter, articulated in a letter to Szilard, still ring true today, ``One can best feel in dealing with living things how primitive physics still is.'' (A. Einstein, letter to L. Szilard quoted in \cite{PS1997}). This is particularly true when dealing with the origin of life itself. It is in the transition from matter to life that our traditional approaches to physics, which accurately describe the predictability of the physical and chemical world, must yield to the novelty and historical-dependency characteristic of life. While much headway has been made in addressing aspects of the puzzle -- ranging from novel explorations of alternative chemistries for life, to self-organization of cooperative networks, to insights gained through application of the tools of statistical physics and thermodynamics -- we have not yet been able to answer the question of how life first emerged. Novel approaches to origins questions that produce a theory of physics that encompasses living matter may ultimately be required to constrain the probability ${\cal P}_{\rm life}$. Constraints could additionally come from astrobiological searches for life. If we are so lucky as to stumble on new fundamental understanding of life that allows us to solve our origins, it could be such a radical departure from what we know now that it might be left to the next generation of physicists to reconcile the unification of life with other domains of physics, as we are now struggling to accomplish with unifying general relativity and quantum theory a century after those theories were first developed.

\section{Acknowledgements} The author wishes to thank Chris Adami and an anonymous reviewer for insightful feedback that greatly improved the content and scope of this review.

\end{document}